# Adaptation and Coevolution on an Emergent Global Competitive Landscape[1]

**Philip Vos Fellman**
Southern New Hampshire University
Manchester, New Hampshire
Shirogitsune99@yahoo.com

**Jonathan Vos Post**
Computer Futures
Altadena, CA
Jvospost2@yahoo.com

**Roxana Wright**
Keene State College
Keene, NH
Rox_wright@yahoo.com

**Usha Dasari**
Human Capital Institute
Austin, TX
Usha_dasari@yahoo.com

[1] The authors gratefully acknowledge the support of Dr. Massood Samii, Chairman, Department of International Business and Dr. Paul Schneiderman, Dean of the School of Business.

# Introduction: Evolutionary Dynamics and Complex Adaptive Landscapes

Notions of Darwinian selection have been implicit in economic theory for at least sixty years. Richard Nelson and Sidney Winter note that while evolutionary thinking was prevalent in prewar economics, the postwar Neoclassical school became almost entirely preoccupied with equilibrium conditions and their mathematical conditions.[2] One of the problems with the economic interpretation of firm selection through competition has been a weak grasp on an incomplete scientific paradigm. As I. F. Price relates in his review, "Organisational Memetics?: Organisational Learning as a Selection Process":[3]

> The biological metaphor has long lurked in the background of management theory largely because the message of "survival of the fittest" (usually wrongly attributed to Charles Darwin rather than Herbert Spencer) provides a seemingly natural model for market competition (e.g. Alchian 1950, Merrell 1984, Henderson 1989, Moore 1993), without seriously challenging the underlying paradigms of what an organisation is. More recently the new physics of chaos and self-organising systems has stimulated some writers, notably Wheatley (1993), to challenge the Newtonian thinking behind the engineering model of the organization…

The difficulty with the "Darwinian" biological metaphor is that in many cases, self-organization may exert a more powerful influence on evolution than selection. In other cases, change occurs not because of selection, but despite selection.[4] In particular, through the use of a quantitative model of "fitness landscapes", Kauffman is able to specify a number of evolutionary characteristics which are quite different than those of the Darwinist or Neo-Darwinist paradigm. The concept of coevolution on a dynamic fitness landscape is perhaps the most important from an economic point of view.

While the concept of the fitness landscape is not new[5] Kauffman was the first to place fitness landscapes in a dynamic setting using Boolean networks.[6] Kauffman explains rugged fitness landscapes as a natural outcome of adaptive evolution:[7]

> Adaptive evolution occurs largely by the successive accumulation of minor variations in phenotype. The simple example of computer programs makes it clear that not all complex systems are graced with the property that a minor change in systemic structure typically leads to minor changes in system behavior. In short, as their internal structure is modified, some systems change behavior relatively smoothly and some relatively radically. Thus we confront the question of whether selective evolution is able to "tune" the structure of complex systems so that they evolve readily,

The mechanism by which biological systems tune is the result of standard statistical mechanics, whereby ensembles move randomly through all phase spaces of the system over time. With respect to the random array of ensemble positions, we might think of economic or business systems which occupy nearby positions in a large market with many securities (a "densely packed" state space). Borrowing from random walk theory we could view the ability of a security to earn economic returns as the measure of its fitness. While this is a very simple metaphor it helps to tie the conceptual foundations of economics and finance to evolutionary biology in the sense that the elements of both systems can rest in peaks or valleys (in finance this would be equivalent to Jensen's $\alpha$, which represents the degree to which management (and by logical inference, corporate structure) either creates or destroys value in a firm. The peaks can be either local maxima or global maxima. In highly efficient markets stocks would reach their equilibrium prices rather rapidly and the fitness landscape would be highly correlated.[8]

---

[2] Nelson, Richard and Winter, Sidney "Evolutionary Theorizing in Economics" Journal of Economic Perspectives, Volume 16, No. 2, Spring 2002
[3] Price, I. F. "Organisational Memetics?: Organisational Learning as a Selection Process", Management Learning, 1995 26: 299-318. Price also notes that "the emergent discipline of complexity theory, most noticeably expressed in the work of the Santa Fe Institute (see Waldrop 1992) and in a recent book by Cohen and Stewart (1994) is revealing the parallels between the processes of self-organisation in economics, physics, biology, and the simulation of artificial life."
[4] See Kauffman, Stuart "The Structure of Rugged Fitness Landscapes" in The Origins of Order, Oxford University Press, 1993. In particular, pp.29-31.
[5] Wright, S. (A) "Evolution in Mendelian populations", *Genetics*, 16 97–159, 1931. (B) "The roles of mutation, inbreeding, crossbreeding and selection in evolution", .Proceedings Sixth International Congress on Genetics 1 356–366 (1932)
[6] Kauffman, S. A. (1969) Metabolic Stability and Epigenesis in Randomly Constructed Genetic Nets. Journal of Theoretical Biology, 22, pp. 437-467.
[7] Ibid. No. 4

In less efficient markets, for example markets like the long distance commodities trades we frequently see in international business, where there are significant information asymmetries and market inefficiencies, the landscape would tend to be "rugged" and multi-peaked. In such a situation, there are opportunities for large arbitrage (similar to the biologist's "punctuated equilibrium").[9] In the context of a dynamic landscape[10] where multiple entities interact and both their presence and their interactions affect the structure of the landscape, Kauffman argues:[11]

> The character of adaptive evolution depends on the structure of such fitness landscapes. Most critically, we shall find that, as the complexity of the system under selection increases, selection is progressively less able to alter the properties of the system. Thus, even in the presence of continuing selection, complex systems tend to remain typical members of the ensemble of possibilities from which they are drawn…Thus, if selection, when operating on complex systems which spontaneously exhibit profound order, is unable to avoid that spontaneous order, that order will *shine through*. In short, this theme, central to our concerns, states that much of the order in organisms may be spontaneous. Rather than reflecting selection's successes, such order may, remarkably reflect selection's failure.

Two important points follow from Kauffman's model. The first is that economic theories particularly equilibrium-oriented theories which model performance as the outcome of selection may have built their structure on an internally contradictory foundation from the ground up.[12] The second point is more technical and has to do with some of the formal mathematical properties NK Boolean networks as well as the mathematical limitations that coevolving populations impose on fitness landscapes. Kauffman's model treats adaptive evolution as a search process.[13]

> Adaptive evolution is a search process—driven by mutation, recombination and selection—on fixed or deforming fitness landscapes, An adapting population flows over the landscape under these forces. The structure of such landscapes, smooth or rugged, governs both the evolvability of populations and the sustained fitness of their members. The structure of fitness landscapes inevitably imposes limitations on adaptive search. On smooth landscapes and a fixed population size and mutation rate, as the complexity of the entities under selection increases an error threshold is reached…the population "melts" from the adaptive peaks and flows across vast reaches of genotype space among near neutral mutants. Conversely on sufficiently rugged landscapes, the evolutionary process becomes trapped in very small regions of genotype space…the results of this chapter suffice to say that selection can be unable to avoid spontaneous order. The limitations on selection arise because of two inexorable complexity catastrophes…each arises where the other does not. One on rugged landscapes, trapping adaptive walks, the other on smooth landscapes where selection becomes too weak to hold adapting populations in small untypical regions of the space of possibilities. Between them, selection is sorely pressed to escape the typical features of the system on which it operates.

Kauffman covers a great deal of ground here, not all of which can be fully recapitulated in a brief paper. Again, Kauffman's Type II complexity catastrophe (which takes place on smooth or highly correlated landscapes) bears a strong resemblance to what Farmer describes as "first order efficiency" in financial markets with very thin arbitrage margins for technical traders.[14] To earn "extraordinary returns" traders are forced to undertake "wide searches", equivalent to highly leveraged positions.

---

[8] For a more complete description see Smith, Eric; Farmer, J. Doyne; Gillemot, Laszlo; and Krishnamurthy, Supriya "Statistical Theory of the Continuous Double Auction", Santa Fe Institute Working Papers, October 20, 2002 http://www.santafe.edu/sfi/publications/Working-Papers/02-10-057.pdf

[9] See Aoki, Masahiko, Subjective Game Models and the Mechanism of Institutional Change, Santa Fe Institute Workshop, http://www.santafe.edu/files/workshops/bowles/aokiIX.PDF

[10] Those unfamiliar with the mechanics of Boolean networks can obtain a very simple, straightforward explanation with examples from Torsten Reil's "An Introduction to Complex Systems", Department of Zoology, Oxford University at http://users.ox.ac.uk/~quee0818/complexity/complexity.html Vladimir Redko provides a slightly more mathematical treatment at http://pespmc1.vub.ac.be/BOOLNETW.html

[11] Ibid. No. 4

[12] See, for example, Reaume, David "Walras, complexity, and Post Walrasian Macroeconomics," Collander David, Ed. <u>Beyond Microfoundations: Post Walrasian Macroeconomics</u>, Cambridge University Press 1996

[13] The search property of organisms seeking maximal fitness levels allows one to draw some additional, strong conclusions about both organisms and landscapes.

[14] Ibid. No. 8, See also Zovko, Ilija and Farmer, J. Doyne, "The Power of Patience: A behavioral regularity in limit order placement", Santa Fe Institute Working Papers, 02-06-27; Daniels, Marcus J.; Iori, Giulia; and Farmer, J. Doyne, "Demand Storage, Liquidity, and Price Volatility" Santa Fe Institute Working Papers, 02-01-001; and Farmer, J. Doyne and Joshi, Shareen, "The Price Dynamics of Common Trading Strategies", in the Journal of Economic Behavior and Organization October 30, 2001.

Kauffman's Type I complexity catastrophe is the result of the fundamental structure of NK Boolean fitness landscapes. In a dynamic fitness landscape, as complexity increases, the heights of accessible peaks fall towards mean fitness. Again, in terms of financial markets we might compare this to the fact that over long periods of time the betas of all stocks in the market have a tendency to drift towards 1.[15]

Bill McKelvey takes the fitness landscape model well past being a metaphor for market evolution and decomposes Michael Porter's value chain into value chain competencies as the parts of firms (the N of the NK network) and K as the connections of between the parts within a firm and the parts of the opponents. He models firm behavior using a heterogeneous[16] stochastic microagent assumption, which then allows him to test four basic concepts: "(1) What intrafirm levels of integrative complexity affect competitive advantage? (2) What levels of integrative complexity influence how rapidly firms in coevolutionary groups reach equilibrium fitness levels, if they do so? (3) What complexity factors might affect the competitive advantage (or height) of fitness levels? and (4) What levels of integrative complexity might affect the overall adaptive success of firms comprising a coevolving system?"[17]

The formal structure of McKelvey's question is every bit as important as the answers, some of which are provided in his work and some which are explained by the Windrum-Birchenhall model of technological diffusion which we discuss next. McKelvey's work serves two important purposes. First, he develops a metric from Porter's value chain which allows him to calculate actual scalar or vector values for different NK structures. This also allows him to calculate fitness levels in terms of Nash Equilibria which vary directly as a function of N and K.[18] Secondly, the values he derives from Porter's value chain allow him to run simulations which specify under what conditions firms should either increase or decrease their internal and external connectivity. While in many ways McKelvey is really just taking the first cut at the pockets. Based on complexity theory, four kinds of complexity are identified. Kauffman's "NK[C] model" is positioned "at the edge of chaos" between complexity driven by "Newtonian" simple rules and rule driven deterministic chaos. Kauffman's insight, which is the basis of the findings in this paper, is that complexity is both a consequence and a cause. Multicoevolutionary complexity in firms is defined by moving natural selection processes inside firms and down to a "parts" level of analysis, in this instance Porter's value chain level, to focus on microstate activities by agents.

---

[15] See Theobald, Michael and Yallup, Peter "Determining the Security Speed of Adjustment Coefficients", http://bss2.bham.ac.uk/business/papers/Theocoeff.pdf , May 2001. See also Blake, David; Lehmann, Bruce N. and Timmerman, Allan "Asset Allocation Dynamics and Pension Fund Performance", November, 1998 http://econ.ucsd.edu/~atimmerm/pensrev2.pdf

[16] See (A) Farmer, J. Doyne "Physicists Attempt to Climb the Ivory Towers of Finance", Computing in Science and Engineering, November-December, 1999 for an explanation of agent-based models. See also W. Brian Arthur's classic game theoretic treatment of an interaction of agents which has no equilibrium and no possibility of satisfying every player either prospectively or retrospectively, the El Farol Bar Game, in (B) "Inductive Reasoning and Bounded Rationality", American Economic Review, May 1994, Vol. 84, No. 2, pp. 406-411. Arthur also deals with the stochastic microagent assumption in(C) Arthur, W. Brian, "Learning and Adaptive Economic Behavior: Designing Economic Agents that Act Like Human Agents: A Behavioral Approach to Bounded Rationality", American Economic Review, Vol. 81, No. 2, May 1991, pp. 353-359.

[17] McKelvey, Bill "Avoiding Complexity Catastrophe in Coevolutionary Pockets: Strategies for Rugged Landscapes", Organization Science, Vol. 10, No. 3, May–June 1999. McKelvey explains his use of Kauffman's model by noting that "The biologist Stuart Kauffman suggests a theory of complexity catastrophe offering universal principles explaining phenomena normally attributed to Darwinian natural selection theory. Kauffman's complexity theory seems to apply equally well to firms in coevolutionary

[18] Ibid., "Kauffman argues that his "NK[C] Boolean game" model behaves like Boolean networks when agent outcomes are limited to 0 or 1, the K number of interdependencies is taken as the number of inputs, and Nash equilibria in N person games are taken as equivalent to agents being trapped on local optima. In the NK Boolean game, fitness yields are assigned to the 0 or 1 actions by drawing from a uniform distribution ranging from 0.0 to 1.0. The K epistatic interdependencies that inhibit fitness yields from an agent's actions are drawn from a fitness table in which fitness levels of each "one-change" nearest neighbor are assigned by drawing from a uniform distribution also ranging from 0.0 to 1.0. Kauffman points out that the complexity tuning effect occurs when increasing K reduces the height of local optima while also increasing their number. Thus, high K leads to complexity catastrophe. In describing how K and C effects enter into the model, Kauffman says:

. . . [F]or each of the N traits in species 2, the model will assign a random fitness between 0.0 and 1.0 for each combination of the K traits internal to species 2, together with all combinations of C traits in species 1. In short, we expand the random fitness table for each trait in species 2 such that the trait looks at its K internal epistatic inputs and also at the C external epistatic inputs from species 1 (Kauffman 1993, p. 244).

One might conclude from this that K and C are combined into one overall moderating effect on the fitness yield from an agent's choice to adopt a higher fitness from a nearest-neighbor. Results of the models indicate otherwise.

As Kauffman points out (pp. 249, 254), the speed at which agents encounter Nash equilibria increases with K, and decreases as C and S increase. Thus, in these models K acts as a complexity "forcing" effect in speeding up the process of reaching stable Nash equilibria, whereas C acts as an "antiforcing" effect, as does S. Presumably K effects are averaged as per the basic NK model, leaving C and S effects (S multiplies the C effects) to modify fitness yields on an agent's actions independently of K effects. The consequence is that increasing K tunes the landscape toward more ruggedness (increased numbers of less fit local optima), and increases the likelihood of agents being marooned on local optima. But increasing C and/or S prevents achieving Nash equilibrium by prolonging the "coupled dancing," as Kauffman (1993, p. 243) calls it, in which opponents keep altering each other's landscapes, keep the fitness search going, and thereby prevent stability—the more opponents there are, the more the instability persists.

problem, he is doing so with rigorous statistical mechanics, so that even where his conclusions are not immediately overwhelming, they are profoundly accurate. Even though we might not expect most firms to start using Kauffman's model in the immediate future (Kauffman did run a joint venture with Ernst and Young for several years, during which his primary purpose was to organize conferences and explain to clients the value and insights of his model)[19] McKelvey's advice, ***In general the simulations indicate that keeping one's internal and external coevolutionary interdependencies just below that of opponents*** can be expected to hold true for quite some time.

Having explored some of the tools of chaos and complexity theory for analyzing evolutionary dynamics, we can now turn to a remarkably compact model developed by Paul Windrum and Chris Birchenhall of the UK, which uses statistical mechanics and computer simulation to predict the detailed impact of technology change including nonlinearities such as network externalities and technology shocks.

---

[19] See Meyer, Chris "What's Under the Hood: A Layman's Guide to the Real Science", Ernst and Young/Cap Gemini Center for Business Innovation, http://www.cbi.cgey.com/events/pubconf/1996-07-19/proceedings/chapter%2010.pdf

# The Windrum-Birchenhall Model of Technological Diffusion by Punctuated Equilibrium with Quasi Homogenous Agent Groups[i]

Windrum and Birchenhall describe their basic model as employing (p. 1):

> …simulation techniques on an agent-based model containing heterogeneous populations of adaptive users and suppliers that co-evolve over time. In the spirit of Nelson and Winter (1982), producers employ various adjustment rules to their production routines, while simultaneously innovating through a combination of imitation and internal R&D. Whereas Nelson and Winter modeled these using a replicator dynamic algorithm, the current paper employs a modified genetic algorithm in which imitation is conducted via a process of selective transfer (one-way crossover) and internal R&D is conducted via selective mutation.

While they may have missed McKelvey's technique of capturing the complete dynamics of heterogeneous agents, their model is otherwise remarkably eclectic, following the fundamental program of Nelson and Winter, incorporating Holland's genetic algorithm, and some approximation of Kauffman's N-1 mutation NK lattice structure. As with Aoki (2000), Windrum and Birchenhall approach technology diffusion and substitution from the standpoint of a punctuated equilibrium. They very carefully define the limits of their model, arguing that:

> Nelson and Winter (1982) argue for engagement in appreciative theorizing as an initial step in the process of developing formal models. One must be clear about what is, and is not, being modeled. As well as identifying a set of stylized facts which the model is expected to replicate – notably, in this case, the ability of a formal model to replicate substitutions by successive species with increasing fitness.[ii]

Similarly, they explain that:

> The basic conceptual and theoretical elements which form the core of the model are drawn from Shy's (1996) discussion of consumer substitution between network size and quality in sequential technology competitions. In contrast to earlier papers by Farrell and Saloner (1985) and Katz and Shapiro (1986), the overlapping generations (OLG) model developed by Shy focuses on repeated technology adoptions. The model allows for different preferences between the 'old' consumer type and the 'young' consumer type, although preferences within each generation are assumed to be identical (i.e. homogeneous).

While this does not provide as heterogeneous a group of agents as either Farmer (2001) or Arthur (2002) employ in their simulations, at least there are two different elements within the population, not unlike the players in Arthur's El Farol bar game. From this basic framework, Windrum and Birchenhall attempt to realize three major goals:

(1) The key question addressed by the model is whether (a) the young consumer type will treat quality and installed user networks as substitutes and, hence, select the later technology (which is de facto assumed to be of higher quality) or (b) alternatively treat them as complements, in which case they will select the old technology.

(2) In this paper we address a number of issues that are 'black boxed' by Shy but which are likely to have important implications for the general thrust of his argument.

   (a) The first issue is that of functional equivalence. A technological substitution involves the substitution of an established product or process by a new alternative that fulfils the same basic function (Grübler et al., 1988; Grübler, 1990).

   (b) Second, there is the quality of alternative designs. The standard economics model of choice ties relative fitness to the\welfare associated alternative commodities. The economic literature contains some important precedents for an analytical treatment of how users first compare, and then rank, substitutable goods (e.g. Frenken, Saviotti and Trommetter, 1999).

   (c) A third issue is the trade-off between quality and price. A trade-off is likely to exist between the quality of the rival technologies and their price, tied to costs of producing these alternative bundles of characteristics. Given that user welfare depends on both the relative performance of each technology and their prices, this will affect demand and hence the outcome of a technological competition (Arthur, Ermoliev and Kaniovski,1987; Arthur, 1989).

   (d) Fourthly, we address is the existence of new types of users. This is likely to be a key factor affecting the complex dynamics of emergent market structures.

Windrum and Birchenhall then argue that "there is likely to be a number of factors influencing user preferences, and that these interact in a complex manner. Three factors in particular were highlighted above:" (P. 4)

  (a)  production costs
  (b)  price
  (c)  performance quality

They formalize this relationship by assigning the following value to the probability of adopting the new technology *B* rather than the established technology *A* at time *t* : (P. 4)

Pr { $U_A(x_A) + v_A(m-p_A) < U_B(x_B) + v_B(m-p_B)$ }
where *x* is the characteristic vector of a technology design
*p* is the price of that design
*V* is the indirect utility of money that can be obtained in other markets

They assume all other markets are fixed, and that the above function has a constant form.

*U(x)* is the direct utility of consuming the good with characteristic vector *x*.

[Note that the utility of not buying a good is *V(m)* and so a user will only accept
offer *(x, p)* if $u_i(x) > V_i(m) - V_i(m-p_i)$] The conclusion which the authors draw, is, in fact, entirely consonant with Arthur, Ermoliev and Kaniovski:

> "In the presence of heterogeneous preferences, a simple analytically tractable solution for equation (1) is unlikely to exist in all but the simplest of circumstances. Multiple equilibria solutions may exist in which it is impossible to predict ex ante whether there will be a technological substitution, a technological lock-out, or a mixed solution (i.e. a partial substitution with market sharing between the old and new technologies). First, multiple equilibria solutions can exist - even when the performance characteristics of one technology application are absolutely superior to those of another - if there is a high frequency of intermediate valuations within the user population. Second, rival technologies typically offer different relative strengths across a set of performance characteristics. Again, given heterogeneous preferences, it is impossible to predict ex ante whether a technological substitution will occur." (P. 4)

Their answer to this quandary is as follows: (p.5)

> "One way of tackling the problem is to construct a simulation model in order to analyze the consequences of heterogeneous user preferences, and the co-evolutionary dynamics of changing user preferences and the innovative activities of competing suppliers."

It is here that Windrum and Birchenhall provide an agent-based iterative model with quasi-heterogeneous agents representing both consumers and firms which co-evolve over time. The model is adjusted so that it brings the groups to market in a random order and gets the groups to determine their demands and purchases. It initiates the replicator dynamic for that period to redistribute the population across user groups, and it gets the suppliers to adjust their capacity, level of production and redesign their goods.

There now follow a welter of simplifying assumptions which lend this model what the authors term "A Keynesian flavor". These assumptions are well within the bounds of good modeling, but they steal much of the potential non-linear "richness" away from the model. On the other hand, they eliminate virtually all of the "noise" of real market data, which allows the authors to carefully study the mutation and co-evolution process, something which might otherwise be obscured by market complexity .[iii] The greatest degree of heterogeneity among agents is lost in the simplification of the pricing model. Its specifications are that: (p. 5)

> "There are M user groups. Associated with group i = 1…M, a utility function $u_i$ is defined over the offer space, namely the Cartesian product X·P of design space X and the price space P (positive real numbers) of the form:
> $U_i(x,p) = \sum_k \alpha_{ik} v_k(x_k) + \beta_i w(m - p) = \alpha_i \cdot v(x) + \beta_i w(m - p)$

> Here $m$ is the budget of the user and is assumed to be the same for all users. The term $\beta_i w(m - p)$ is the indirect utility obtained by spending the residual budget in other markets. All users in the same group are assumed to adopt the same utility function. Each supplier offers to sell a good with some design $x$ at some price $p$. Users use these utility functions to rank alternative offers and as a measure of well-being. Note that users always have the option of not accepting any of the offers and may keep all of their budget for use elsewhere. The utility of this option is $\beta_i w(m)$ and will be called the null utility. It can be seen that the utility functions differ across groups only in having different values for the coefficients $\alpha_i$ and $\beta_i$."

Another simplification is the component function: (p. 5)

> "Currently we use a simple square root function for the component functions, i.e.
>
> $$v_k(x_k) = \sqrt{x_k} \quad \text{and} \quad w(\tilde{m}-p) = \sqrt{(\tilde{m}-p)}$$
>
> The population of users in each period is G and a form of the replicator dynamics described below governs the distribution across the M groups. Let $G_{it}$ be the number of users of type $i$ at time $t$. We use the subscript $t$ only when necessary to distinguish between periods. In each period supplier $j$ offers a quantity $Q_j$ of a particular design-price combination $(x_j, p_j)$. After suppliers have 'posted' these offers, user groups appear in the market in a random order. Let I(i), with i=1,..,M, be a permutation of the indices {1,.., M} so that I(1) is the first group to come to market. Note that this permutation will differ from period to period. Given the utility function $U_{I(1)}$ associated with this group, the users rank the offers $(x_j, p_j)$ in descending order of preference.
>
> Let J(j) j = 0,1,..,M represent this ranking, so that J(j) is a permutation of {0, 1,..,M}, where 0 represents the 'null offer', i.e. buy none of the goods. If the null offer (J(0) = 0) is best, the users in that group exit the market without buying anything. If the supplier ranked highest by the users has an offer which dominates the null offer (i.e. J(0) > 0) then all users in that group will 'post' a demand for one unit of that offer. If supplier J(0) has produced a sufficient quantity of the good (i.e. $Q_{J(0)} \geq G_{I(1)}$) then all these demands will be converted into sales, all users in the group exit the market and the available quantity of the good is reduced by the volume of sales, i.e. $Q_{J(0)} \Leftarrow Q_{J(0)} - G_{I(1)}$    (p. 6)

> "If demand exceeds supply (i.e. $Q_{J(0)} < G_{I(1)}$) then $Q_{J(0)}$ demands are converted into sales, $Q_{J(0)}$ of the user leave the market and the available quantity of the good becomes zero and the remaining users $G_1 - Q_{J(0)}$ consider their next best option J(1). The interaction of these remaining users with this offer is identical to interaction with J(0). If J(1) = 0 they leave the market, otherwise they post demands for the goods and these are met fully or partly depending on the quantity $Q_{J(1)}$ on offer. This process for group I(1) continues until all users in the group have left the market. Group I(2) enters the market and interacts with suppliers in the same way apart from the fact that the quantities available to this group will be reduced by any sales made to group I(1) users. This continues until all groups have entered and left the market. When group I(i), i > 1, enters the market the quantities available will be the $Q_j$'s minus any sales made to user groups I(k) for k=1,.., i-1." (p. 6)

> "After this process in period t each user group will have attained an average level of utility $W_{it}$. This is the average utility of the users in the group after they have consumed any good bought in this market. Note that all users will attain a utility no less than the null utility and thus $W_{it}$ will be no less than the null utility." (p. 6)

The preceding step is a nice simplification on the part of the authors, guaranteeing (albeit a tad unrealistically) that no user sacrifices utility. They next incorporate the parameters for the situation when demand exceeds supply:

> "If demand exceeds supply (i.e. $Q_{J(0)} < G_{I(1)}$) then $Q_{J(0)}$ demands are converted into sales, $Q_{J(0)}$ of the user leave the market and the available quantity of the good becomes zero and the remaining users $G_1 - Q_{J(0)}$ consider their next best option J(1). The interaction of these remaining users with this offer is identical to interaction with J(0). If J(1) = 0 they leave the market, otherwise they post demands for the goods and these are met fully or partly depending on the quantity $Q_{J(1)}$ on offer. This process for group I(1) continues until all users in the group have left the market. Group I(2) enters the market and interacts with suppliers in the same way apart from the fact that the quantities available to this group will be reduced by any sales made to group I(1) users. This continues until all groups have entered and left the market. When group I(i), i > 1, enters the market the quantities available will be the $Q_j$'s minus any sales made to user groups I(k) for k=1,.., i-1. After this process in period t each user group will have attained an average level of utility $W_{it}$. This is the average utility of the users in the group after they have consumed any good bought in this market. Note that all users will attain a utility no less than the null utility and thus $W_{it}$ will be no less than the null utility."    (p. 6)

The authors then calculate the new distribution for the next iteration as follows:

> "Let $\rho_{it} = \frac{G_{it}}{G}$, where G is the total population, be the proportion of the user population in the $i^{th}$ group. Given these utilities, a new distribution, $\rho_{it+1}$ is calculated as :
>
> $$\rho_{it+1} = \frac{\rho_{it} r W_{it}}{\sum_k \rho_{kt} r W_{kt}}$$
>
> Where $r$ is the factor determining the strength of the replicator effect of the differing utilities. Groups with above average utilities grow larger and groups with below average utility decline, i.e., they have a negative growth rate." (p. 6)

The authors work out similar dynamics for suppliers, including prices, profits, wealth and capacity. While some of these parameters are rather simplistic, the model almost certainly gains more in clarity of exposition than it loses in relationship to real market complexity. On the whole, the sum of these simplifications and linearization is, in the authors' own words to produce "a model of production that is distinctly Keynesian in flavor as adjustments to excess demand occur primarily through changes in output and production capacity rather than price." Having hopefully addressed any further objections, it is interesting to note that Japan's economy recovery in 1932-36 was based on the Keynesian macro-policies of Finance Minister Takahashi Korekiyo, and that in looking at institutional evolution with punctuated equilibrium, a Keynesian model may offer some distinctly Japanese advantages.[iv]

The most interesting part of this model is not, however, the elegant construction of supply and demand or even the welfare function. What is truly a masterful achievement here is that because the variation among some rather ordinary (and one would hope, rather well understood) features of the model have been artificially dampened, the process of innovation, in this case modeled with an increasing fitness replacement function allows the authors to demonstrate (much in the fashion of Arthur, Ermoliev and Kaniovski demonstrating large differences in outcome arising from small stochastic perturbations of the system) the powerful effects of a punctuated equilibrium on overall fitness, even though any individual change is both small and random. The mutation process is executed in two stages. This facilitates the "adaptation" process. During the first stage, each of the suppliers considers mutations:

> "Mutations are carried out in isolation of other suppliers. Given design $x$ for the $j^{th}$ supplier period t, the supplier considers a mutated design $x^*$. Each component $x^*_i$ mutates with probability $\mu$ and if it does mutate it has the value $x_i + \kappa\varepsilon$, where $\kappa$ is a mutation factor and $\varepsilon$ is a random number drawn from a standard normal distribution. The mutated design replaces the current design only if it increases the utility of the supplier's target user type. After mutation, suppliers consider further innovation based on imitation of rival suppliers. Each supplier picks another supplier in a biased random draw from the existing set of suppliers. This selection is biased toward to the more profitable suppliers. In fact it is based on Goldberg's 'roulette wheel' in that the probability of supplier $j$ being selected is proportional to the profit made by the supplier $j$ in the current period. Having selected a rival, the supplier creates a new candidate design $x^*$ by transferring part of the rival's design $x^r$ to replace the matching elements of its current design. A random set H of characteristics is selected, as shown:[v]
>
> $x_h^* = x_h^r$ for $\in H$ and $x^*_h$ and $x_h^* = x_h$ for $\notin H$.

This selective transfer operator is different to crossover in genetic algorithms. Here there is no mutual exchange of elements, selective transfer is a one-way emulation. Hence the supplier that is being emulated does not have to adjust its design as a consequence of this operator. The new design x* replaces the current design x only if this increases the utility of the target user type." (p. 8)The authors then introduce a technology shock into the system:

> "There is a technological shock at period $T_1$. This shock has three features. First, the characteristic space qualitatively changes, i.e. the set of characteristics associated with the new technology application differs from that of the old technology. More specifically, prior to time $T_1$, the characteristic space has characteristics dimensions 1 to $h_1$. After $T_1$, the characteristic has dimensions $h_2$ to h, where $1 \leq h_2 \leq h_1 < h$ Furthermore, before $T_1$ there is a limit $x_{max}$ on characteristic values. We use $D_1$ and $D_2$ to represent the design spaces before and after $T_1$ respectively. Before $T_1$ all designs must belong to $D_1$ and users place positive weight on IS application characteristics 1 to $h_1$ and zero weights on characteristics $h_1+1$ to h. Second, whereas suppliers prior to $T_1$ are producers of the old technology, after $T_1$ all market entrants are 'new' technology producers. At the same time, the user groups that emerge after $T_1$ are 'new' technology users. New generations of users and suppliers are generated in the following way; at $T_1$ 'dead' suppliers and 'dead' user types are replaced by new generations of suppliers and user types. A supplier is treated as 'dead' if its market share has fallen below a cut-off value and a user type is 'dead' if its share of the user population has fallen below a cut-off value. The new suppliers created at time $T_1$ must provide designs in $D_2$. New user types place positive weight on IS application characteristics $h_2$ to h and zero weight on characteristics 1 to $h_1$-1. Third, picking up on the earlier discussion of the possible importance of the relative rate of falling unit costs due to static and dynamic economies of scale, the cost of production for new technologies is reduced by a factor θ, i.e. after $T_1$ all $γ_k$ are reduced by a factor θ." (p. 9)

Using a quasi-random number generator for each successive run, organized by a batch control process, they proceed to test whether the model can actually produce adaptive results with an overall higher fitness level. They argue that "we believe overall…the model does indeed produce the patterns associated with technological substitutions within a market niche. The envelope of aggregate utility is raised as the new technology displaces the old technology in the market niche. A substitution occurs when a new, fitter technology (i.e., one offering higher levels of welfare) displaces and older, less fit technology (i.e., one offering lower levels of welfare)." (p. 9)

While these results shouldn't really surprise anyone who understand the values and constraints imposed by the rules (perhaps the proper term here is "meta-strategies" or "meta-behaviors"), there were two very interesting results from this study, which equilibrium economic models might not capture well, if at all. In one case a new technology replaced an old technology and the user base was also replaced with new users. One might compare this process to the substitution of entertainment professionals, "stars", or to the shift from Vinyl and tape to CD, or even the interesting example of Encyclopedia Britannica being replaced by Microsoft public domain content software (Evans and Wurster, 1997).

The second set of novel results found new technologies replacing older technologies, but this time the technology went to the established user base. While a complete explanation for this phenomenon is not provided, the authors appear to have tested their variables, as well as the substitutability of variables as far as they could, with the considerable resources at their disposal. They then conclude that what drives the model is the replicator dynamics and the utility of the targeted group. While co-evolution, or simultaneous deformation of the fitness landscape is not undertaken (in that sense this is a static, multiple iteration model rather than a dynamical systems model) the authors do raise some interesting issues about the potential value of technology change when they conclude, "Type A substitution tends to occur if a new Type Two firm can quickly generate design and price combination that would make its targeted Type Two customers better off than the current dominant Type One firm-consumer alliance." (p. 11)

The implications for industrial policy as well as corporate strategy are that first mover advantages play a crucial role in technology substitution, but that they are "emergent" and not discernible below a certain threshold of market penetration (Modis, 1998). However, both the Windrum Birchenhall technology replacement model and the combinatorial market landscape approach suggest that markets and consumers need to be "tolerant" of new product entrants or new product categories if they expect to raise overall welfare, particularly if the expected mechanism of national prosperity is characterized as a punctuated equilibrium Aoki, (1997, 1998C). These findings, curiously enough, have a bit of the flavor of the "Chicago School" treatment of consumption and production insofar as they argue that value clusters are built from the ground up, often using local rules of behavior (Holland, 1995, Kauffman 1995, Aoki, 1995B).

Similarly, this line of analysis suggests that while new technologies may offer potentially high returns to scale and scope, the operating environment at the firm level needs to be flexible and capable of rapid adaptation in order to capture the gains which a high degree of innovation can provide (Lissack, 1996, 2000). In this sense, one might argue generically that in order to optimize returns, the organization must imitate the product. Michael Porter argues convincingly that for many industries, corporate strategy needs a time-line of a decade or more in order to establish a unique strategic position based on internal cross subsidization (Porter, 1997). However, in fields where the product life cycle runs from six to eighteen months (as it does in the case of microprocessors and many other high technology, high value-added products) a different kind of model than the traditional "five forces" or "value cluster" model is required (Slater and Olson, 2002).

However, the mid-twentieth century dynamics of the Chicago school (if one can properly characterize any of this work as "dynamics" at all) are insufficient to characterize the relationship between government and regulation to the production function of the firm. Indeed this kind of analysis is far better suited to treatments of relatively homogenous consumer preferences, which can then be aggregated as an overall welfare function.

On the other hand, the policy message of the non-linear models discussed above (primarily the NK fitness model, as applied through evolutionary walks on a hypercube of competitive product and state spaces as represented by replicator dynamics) argues that if government policy doesn't support superior technologies and leading edge sectors, then the country's entire technology base may ultimately be at risk in the long run. This conclusion is a relatively strong rejection of the efficient market hypothesis, insofar as the empirical level demonstrates rather large information asymmetries across buyers, sellers, consumers, producers and inventors in the area of new technology.

At the firm level, management in general and top management in particular, needs to exercise an especially sensitive degree of restraint before sidelining or dropping new technology products (Modis 1998). Another broad recommendation which comes out of this research is that to be competitive in future markets, large organizations, including large multinational corporations will need to become less hierarchical, less vertically integrated and to adopt more flexible organizational structures if they are to avoid becoming trapped on local fitness landscape maxima. In recent years, a new subfield of organizational behavior, perhaps best described as "organizational complexity" which focuses on just such problems.

Again, one of the most important conclusions of this group of researchers is that in the 21$^{st}$ century, organizations will need to decentralize and move away from more traditional divisional structures and hierarchical control if they are to survive on the emerging technology landscape (Slater and Olson, 1992). A number of new techniques have been designed to address this problem, including "simulated annealing", "organizational patches" and NK fitness measures of organizational interconnectedness (Lissack, McKelvey, 1999, Rivkin and Siggelkow, 2001, and Frenken and Valente, 2002). Firms which cannot adapt their internal structure to the new dynamics of technological competition are not likely to prosper, nor indeed survive in the turbulent years to come (Evans and Wurster, 1997).

# Appendix: Internal Structure of the Genetic Algorithm – Jonathan Vos Post

One thing I found, well ahead of Koza and other researchers, through my experiments in 1975-1977, at the University of Massachusetts at Amherst, where I beta-tested John Holland's book "Complexity in Natural and Artificial Systems" by coding the Genetic Algorithm into APL and running evolution of software was as follows:

***The evolving software must implicitly determine HOW rugged the fitness landscape is, and adapt its mutation rate, cross-over rate, and inversion rate accordingly.***

My insight beyond that notion was to explicitly add genes that set mutation rate, cross-over rate, and inversion rate right at the end of (as it turned out, location didn't much matter) the existing genes. That is, I put **meta-variables** which coded for parameters of the Genetic Algorithm itself in among the variables coded for the underlying evolving system.

I then ran the Genetic Algorithm with the extended "chromosomes" on landscapes of different ruggedness. As I'd hoped, the 3 types of mutation rates themselves evolved to good rates that fit the rates optimum for adaptation on that degree and style of ruggedness.

This proved, at least to my satisfaction, that the Genetic Algorithm was not only parallel as explicitly obvious, and parallel in its operation on the higher-dimensional superspace of possible gene-sequences, as Holland demonstrated (in his example where a binary string of 0-1 alleles was part of the subspace of trinary strings of 0-1-"don't care" so that spaces of dimension 2-to-the-power-of-N in which evolution occurred were faithfully sampling and implicitly evolving in superspaces of dimension 3-to-the-power-of-N, where N is the bit-length of the evolving "chromosome"), but parallel at yet a higher level, namely that *the Genetic Algorithm worked at simultaneously evolving the parameters of its own operation in rugged landscapes while evolving the simulated organisms that had their fitness determined by that same landscape.* My analogy was to "hypervariable" genes in the immune system.

With the infinitely greater computational power and speeds available today, this experiment should, in principle, be repeatable in a multi-generational context which should then allow for the testing of emergent order for global evaluation functions. Until now, evolutionary simulations have been based either on the genetic algorithm[20] or similar structures or else using assumptions that finesse the questions of how global evaluation emerges at all and simply proceeds to use global criteria in a purely local environment.

---

[20] Holland, John and Mimnaugh, Heather, <u>Hidden Order</u>, Perseus Publishing, 1996

# END NOTES

[i] This treatment was drawn from Windrum, P., Birchenhall, C., 'Modeling technological successions in the presence of network externalities', Danish Research Unit for Industrial Dynamics Conference in Honour of Richard Nelson and Sydney Winter, Aalborg, 12th - 15th June 2001.

[ii] The Authors cite Nicolis and Prigogine (1977), noting that "In their discussion of species substitutions in ecological systems, Nicolis and Prigogine add that the new invaders must have a better capability of exploiting the same resources offered within an ecological niche. In other words, they must be able to do something 'more' or 'better' - whether it be capturing a certain type of prey, reproducing or avoiding death - than the previous incumbent. As a consequence, the fitness of successive species occupying a given niche will increase over time. The envelope of overall fitness is raised as more efficient species displace earlier incumbents within an ecological niche." (p. 3)

[iii] Farmer (2001) notes with respect to Agent-Based Modeling: "So far as I know, no one currently uses agent based models for investment. In the context of this presentation, an agent-based model involves a
model for price formation and a model for agent behavior, including the models' information inputs
and how the inputs behave, which could involve learning. From an investor's point of view, agent based
models represent a new direction that may or may not result in more practical investment models.  From an academic's point of view, agent-based models are fascinating new tools that make it possible to investigate problems previously out of reach… My own experience at Prediction Company and elsewhere illustrates the difference between agent-based models, which are a fundamental approach to modeling, and empirical  models, which are a strictly practical, data-driven approach.1 Norman Packard, Jim McGill, and I were the senior founding partners of Prediction Company. When we founded Prediction Company, we did not know anything about financial markets. We were experts in time-series modeling, which is an empirical approach to building models for data that are ordered in time. I came to be involved in time-series forecasting because of my earlier experience in the field of chaotic dynamics. I developed nonlinear time-series models that could take advantage of low-dimensional chaos and, in some cases, give improved forecasts for such problems as fluid turbulence, ice ages, and sunspot cycles.  Prediction Company uses a purely empirical approach to market forecasting. The models are not based on a fundamental, reasoned understanding of how markets work; instead, they are based on a search for patterns in historical data. The fundamental premise is that patterns that existed in the past will repeat themselves in the future. These patterns are usually rather subtle. The market is pretty efficient in that extremely few persistent simple patterns exist. It is not quite so efficient with respect to complex patterns, and if you look carefully and comprehensively, you may find things that very few have seen before.  The big problem is that once you allow for complex patterns, the number of possibilities is enormous and careful statistical testing becomes essential. Feature extraction is a key aspect of Prediction Company's models. In the extraordinarily noisy environment of the financial world, statistical modeling methods that blindly generate models directly from data do not work well. The problems are too much noise, the limited size of datasets, and the fact that techniques have a hard time finding patterns that are real and persistent rather than ephemeral random fluctuations. Throwing away distracting information is crucial…The key to good agent-based modeling is to capture the agent behaviors as realistically as possible with a minimum of assumptions. Unfortunately, in many people's minds, agent-based modeling has come to be associated with highly complicated models in which each agent has his own customized kitchen sink. Such models are so complicated that they are almost as hard to understand as real markets.  In order to enhance understanding, and to ever have a hope of using models for prediction (and fitting their parameters to real data), it is important to keep agent-based models simple…. Real traders have credit limits (i.e., there is a wealth-dependent limit to the size of the positions they can take). If a trader tries to take a position that is too big with respect to her credit limit and wealth, her position will be capped. If the agent's wealth decreases, she may have to sell some assets to stay within the limit. This limit introduces a nonlinearity into the agent's strategies that has several effects on the price dynamics of the market. It can cause prices to deviate even more from fundamental values and, in extreme cases, can even make prices oscillate on their own.   Another key element is reinvestment. As traders make profits, they change the scale of their trading accordingly. This behavior means that the c parameters are not fixed;\ rather, they are functions that depend on wealth and credit limits. When we put all of these elements together, we find that the market does indeed tend toward the point where everything is efficient. This point is somewhat noisy; there are fluctuations around it, but each agent's wealth stays more or less fixed. At this
point, there is equilibrium among the market maker, value investor, and trend follower.

[iv] See Morley, James ed., Economic Dilemmas of Prewar Japan, Yale University Press (New Haven: 1972)

[v] Chris Meyer (1996) explain this kind of evolutionary mechanism in terms of a kitchen chair with variable features, seat, back, arm-rest, etc. which can be made with either wood or plastic and which have a higher fitness if one material is used rather than another.  In order to explain how changing the configuration, or mutation improves fitness and how small random walks can produce large changes in position, he begins with a Boolean hypercube as an evolutionary fitness mapping mechanism.  As Chris Meyer explains "The majority of Kauffman's talk centered on the mathematical models used by complexity scientists to depict what biologists call "fitness landscapes." "Adaptation is usually thought of as a process of 'hill climbing' through minor variations toward 'peaks' of high fitness on a fitness landscape," Kauffman wrote in At Home in the Universe. "And natural selection is thought of as 'pulling' an adapting population toward such peaks.  We can imagine a mountain range on which populations of organisms (or in this case, programs) are feeling their way to the summits.  Depending on whether it is beneficial, a random change in the genome (the computer code) puts a mutant higher or lower on the terrain. The search problem is compounded by the fact that the evolving population cannot actually see the contours of the landscape. There is no way to soar above and see a God's eye view.  We can think of the population as sending out 'feelers' by generating, at random, various mutations. If a mutation occupies a position higher on the terrain, it is fitter, and the population is pulled to the new position." In his lecture, Kauffman described two fitness landscape models in detail: the Boolean hypercube and the NK model. "While these models are mere beginnings, they already yield clues about the power and limits of natural selection," he explained in his book. In his lecture, he said, "I'm going to show you how they give us the tools to think about such features in economic systems as learning curves of the kind that Michael Rothschild was talking about yesterday. In short, I think we can think about a lot of things about technological innovation, just based on the notion of a hard combinatorial optimization problem."  To introduce the concept of a Boolean hypercube, he said, "let's think about a chair made of four parts. It has a back, seat, legs, and arms. And suppose that the parts come in two flavors, say, wood or plastic." That means there are 24—or 16—possible kind of chairs, depending upon whether the back is wood or plastic, the seat is wood or plastic, the legs are wood or plastic, and the arms are wood or plastic. If you Figure look at a Boolean Hypercube, you see that each of these chair part options can be assigned to a

vertex on the Boolean hypercube. Kauffman then introduced two concepts associated with the Boolean hypercube fitness landscape: combinatorial objects and one-mutant neighbors. All of the options you present on a Boolean hypercube are defined as combinatorial objects because they show the symbolic relationships between discrete mathematical elements belonging to a finite set. The concept of a one-mutant neighbor is used to evaluate whether changing between two associated options on the landscape—changing the chair back from wood to plastic, for example—would make the end result "fitter." Boolean hypercubes are set up so that if you first located the wood chair back, its alternative "mutant," the plastic chair back, would be found on a neighboring vertex. Fit and Functional Effectiveness "Whether I'm talking about peptides or chairs, I need a notion of the fit and the functional effectiveness of the entity," Kauffman explained. "Suppose I think of the affinity of a peptide for the estrogen receptor as the fitness of that peptide for binding the estrogen receptor. Or I think of the comfort of the chair. . . the fitness of the chair for my derriere." The simplest way to assess the fitness of the fictional chair is to rank order the options, from the worst chair to the best chair. In Kauffman's scale, one was the worst and 16 was the best. Kauffman assigned the numbers randomly. Once the fitness of each chair option was established, Kauffman took us on an "adaptive walk." The term aptly describes the process complexity scientists use to evaluate the fitness of alternative options—in this case, chair comfort—on a fitness landscape. "I start somewhere—for example, at the worst chair or peptide—and I look in every direction and the rule of the game is, if my neighbor is better than me, I get to go there. So, I'm going to put an arrow pointing in the direction that is uphill, up the fitness landscape, where you're trying to go up to get better. And that orients all of these edges with arrows." "The simplest thing you could do is start somewhere and follow the arrows from tail to head and what happens is, pretty soon you get to some spot where you can't go up any further," he continued. "You're at a local peak on the landscape." There were three local peaks on our chair fitness landscape. It's important to know how many peaks there are when you're evaluating a fitness landscape, as well as how many steps it takes to climb up to them. "You don't have to be a rocket scientist to realize that as you go uphill, the number of directions uphill dwindles somehow. To say that the directions uphill dwindle means that as you go uphill, it's getting harder and harder and harder to find a way to improve," Kauffman explained. "And that's going to underlie big chunks of learning curves." Real World Landscapes Aren't Random The problem with the first fitness landscape Kauffman demonstrated using the Boolean hypercube, he explained, "is that it's the simplest case, because it's the hardest case: it's a random landscape. In the real world, nearby points typically are similar heights. Landscapes aren't random." In At Home in the Universe, he writes: "No complex entity that has evolved has done so on a random fitness landscape. Things capable of evolving—metabolic webs of molecules, single cells, multicellular organisms, ecosystems, economic systems, people—all live and evolve on landscapes that themselves have a special property: they allow evolution to 'work.' These real fitness landscapes, the types that underlie Darwin's gradualism, are 'correlated.' Nearby points tend to have similar heights. The high points are easier to find, for the terrain offers clues about the best direction to proceed."